\documentclass[12pt]{article}

\usepackage{amsmath}
\usepackage{amssymb}
\usepackage{amsthm}
\usepackage{verbatim}

\usepackage{tikz,pgfplots}
\usetikzlibrary{plotmarks} 

\usepackage{graphicx}

\usepackage{hyperref}

\usepackage[lined,algoruled,noline]{algorithm2e}

\newcommand{\bY}{\boldsymbol{Y}}
\newcommand{\by}{\boldsymbol{y}}
\newcommand{\bX}{\boldsymbol{X}}
\newcommand{\bx}{\boldsymbol{x}}
\newcommand{\bZ}{\boldsymbol{Z}}

\newcommand{\bW}{\boldsymbol{W}}
\newcommand{\btheta}{\boldsymbol{\theta}}
\newcommand{\bepsilon}{\boldsymbol{\epsilon}}
\newcommand{\bA}{\boldsymbol{A}}
\newcommand{\ba}{\boldsymbol{a}}
\newcommand{\bv}{\boldsymbol{v}}
\newcommand{\be}{\boldsymbol{e}}
\newcommand{\bmu}{\boldsymbol{\mu}}
\newcommand{\bB}{\boldsymbol{B}}
\newcommand{\bL}{\boldsymbol{L}}
\newcommand{\bH}{\boldsymbol{H}}
\newcommand{\bh}{\boldsymbol{h}}
\newcommand{\bU}{\boldsymbol{U}}
\newcommand{\bu}{\boldsymbol{u}}
\newcommand{\bT}{\boldsymbol{T}}
\newcommand{\bzero}{\boldsymbol{0}}
\newcommand{\bI}{\boldsymbol{I}}
\newcommand{\bSigma}{\boldsymbol{\Sigma}}

\newcommand{\bsy}[1]{\boldsymbol{#1}}

\newcommand{\Nor}{\mathrm{N}}

\usepackage{xcolor}

\usepackage{soul}

\title{Bayesian Linear Models: A compact general set of results.}

\author{
	J. Andr\'es Christen\\
	Centro de Investigaci\'on en Matem\'aticas (CIMAT), CONACYT,\\
	Guanajuato, GTO, Mexico, \textit{jac@cimat.mx}
}

\date{Nov 2023}

\begin{document}

\bibliographystyle{plain}

\maketitle

\begin{abstract}
I present all the details in calculating the posterior distribution of the conjugate Normal-Gamma prior in Bayesian Linear Models (BLM), including correlated observations, prediction, model selection and comments on efficient numeric implementations.  A Python implementation is also presented.  These have been presented and available in many books and texts but, I believe, a general compact and simple presentation is always welcome and not always simple to find.
Since correlated observations are also included, these results may also be useful for time series analysis and spacial statistics.  Other particular cases presented include regression, Gaussian processes and Bayesian Dynamic Models
\end{abstract}

\section{Preliminaries}



These notes are \textit{not} intended to explain the details of Bayesian regression models \cite{broemeling2017}, fitting Gaussian processes \cite{rasmussen2006} \cite{ohagan1994}, Bayesian time series (dynamic models) \cite{west1997} and Bayessian linear models in general.  The cited books should serve as a starting point for learning about each topic.  However, many  of the models involved in those topics, and others related, are based on the same basic calculations, namely, the derivation of the Normal-Gamma conjugate posterior distribution, the Normal and t predictive and the normalization constants for model comparisons, in the Bayesian analysis of linear models.  This paper presents, in detail, the basic algebra involved in such derivations.
 While these derivations have been presented and are available in numerous books and texts, as previously mentioned, I believe that a concise, general, and accessible presentation is always valuable and often not easy to find.
Moreover, I include some comments on the efficient numeric evaluation for the required hyper parameters and a compact Python class with an implementation \url{https://github.com/andreschristen/BLM}.  I also include some examples as an illustration.

I include some preliminary results that will be required:

\begin{itemize}
\item The Sherman-Woodbury-Morrison lemma \footnote{\url{https://en.wikipedia.org/wiki/Woodbury_matrix_identity}}.  Let $\bsy{E}$ be a $n\times n$ and $\bsy{D}$ a $k\times k$ invertible matrices and $\bB$ $n\times k$ and $\bsy{C}$ $k\times n$ matrices.  Then
\begin{equation}\label{eqn:SWM_lemma}
(\bsy{E} + \bB \bsy{D} \bsy{C} )^{-1} = \bsy{E}^{-1} - \bsy{E}^{-1}\bB( \bsy{D}^{-1} + \bsy{C} \bsy{E}^{-1} \bB)^{-1} \bsy{C} \bsy{E}^{-1}.
\end{equation}

\item The matrix determinant lemma\footnote{\url{https://en.wikipedia.org/wiki/Matrix_determinant_lemma}}. Let $\bsy{E}$ be a $n\times n$  invertible matrix and $\bu$ and $\bv$ $n\times 1$ vectors, then
\begin{equation}\label{eqn:MD_lemma}
\left| \bsy{E} + \bu \bv' \right| =
\left| \bsy{E} \right| (1 + \bv' \bsy{E}^{-1} \bu).
\end{equation}

\item The multivariate Normal distribution\footnote{\url{https://en.wikipedia.org/wiki/Multivariate_normal_distribution}} \cite{eaton2007}.  If the ($n \times 1$ column) vector $\bY \sim \Nor_n ( \bmu, \bSigma)$ then:

\begin{enumerate}
\item  The pdf of $\bY$ is
$$
f(\by) = (2 \pi)^{-\frac{n}{2}} \left| \bSigma \right|^{-\frac{1}{2}}
\exp\left\{ -\frac{1}{2} (\by - \bmu)' \bSigma^{-1} (\by - \bmu) \right\} . 
$$
$V(\bY) = \bSigma$ is the positive definite covariance matrix and $E(\bY) = \bmu$ the mean vector.

\item \label{property3NM} Marginally, each entry of $\bY$ is univariate normal $Y_i \sim \Nor( \mu_i, \sigma_{i,i})$.
\item If $\bY = [ \bY_1, \bY_2 ]'$, with $\bY_1$ and $\bY_2$ being $q \times 1$ and $(n-q) \times 1$ subvectors, respectively, the conditional distribution is
$$
\bY_1 \mid \bY_2=\by_2 \sim \Nor_q(
\bmu_1 + \bSigma_{12} \bSigma_{22}^{-1} (\by_2 - \bmu_2),
\bSigma_{11} - \bSigma_{12} \bSigma_{22}^{-1} \bSigma_{12} 
) .
$$
Here, the covariance matrix $\bSigma$ is partitioned conformably with $\bY$ as
\[
\bSigma =
\begin{pmatrix}
\bSigma_{11} & \bSigma_{12} \\
\bSigma_{21} & \bSigma_{22}
\end{pmatrix},
\]
where $\bSigma_{11}$ and $\bSigma_{22}$ are the covariance matrices of $\bY_1$ and $\bY_2$, respectively, and $\bSigma_{12} = \bSigma_{21}'$ represents the cross-covariance between $\bY_1$ and $\bY_2$.
\item If $\bB$ is a $q \times n$ matrix and $\ba$ a $q \times 1$ vector then
$$
\bB \bY  + \ba  \sim \Nor_q ( \bB \bmu  + \ba , \bB \bSigma \bB').
$$
\end{enumerate}

\item  For the multivariate t distribution\footnote{\url{https://en.wikipedia.org/wiki/Multivariate_t-distribution}}. If ($n \times 1$ column) vector $\boldsymbol{\Lambda} \sim T_n^\nu (  \bmu, \bsy{D})$ then its pdf is
$$
f(\boldsymbol{\lambda}) =
\pi^{-\frac{n}{2}} \left| \nu \bsy{D} \right|^{-\frac{1}{2}} \frac{\Gamma\left( \frac{\nu + n}{2}\right) }{\Gamma \left( \frac{\nu}{2}\right)}
\left[ 1 + \frac{(\boldsymbol{\lambda} - \bmu)' \bsy{D}^{-1} (\boldsymbol{\lambda} - \bmu) }{\nu}\right]^{-\left(\frac{\nu + n}{2}\right)},
$$

$\nu$ are the degrees of freedom, $E(\boldsymbol{\Lambda}) = \bmu$ is the mean vector, $\bsy{D}$ is the dispersion matrix, and $V(\boldsymbol{\Lambda}) = \frac{\nu}{\nu-2} \bsy{D}$ if $\nu > 2$, otherwise $\boldsymbol{\Lambda}$ does not have second moments.

\item If $Y \sim Ga(  \alpha, \beta)$ (the Gamma distribution), its pdf, in our preferred parametrization, is
$$
f(y) = \frac{\beta^\alpha}{\Gamma(\alpha)}
y^{\alpha-1} \exp\{ -y \beta\} I_{(0,\infty)}(y).
$$
$\alpha$ is the shape parameter and $\beta$ is the \textit{rate} parameter.  Here $E(Y) = \frac{\alpha}{\beta}$ and $V(Y) = \frac{\alpha}{\beta^2}$.

\item Finaly, from the Gamma pdf one can see that
$$
\int_0^\infty y^{\alpha-1} \exp\{ -y \beta\} dy =
\beta^{-\alpha}\Gamma(\alpha) .
$$

\end{itemize}

\section{The model, the prior and the posterior}

The linear model is
\begin{equation}\label{eqn:reg}
\bY = \bX \btheta + \bepsilon;
\bepsilon \sim \Nor_n( \bzero, \lambda^{-1} \bSigma).
\end{equation}
Where:
\begin{itemize}
	\item $\bY$ is a $n \times 1$ vector of responses.
	\item $\bX$ is a $n \times p$ design or covariate matrix
	
	\item $\btheta$ is a $p \times 1$ vector of unknown parameters.
	\item $\bSigma$ is a $n \times n$ variance-covariance matrix, and will be taken as known.  The general case of unknown covariance is not considered here. We let $\bA = \bSigma^{-1}$, the  $n \times n$ precision matrix.
	\item $\lambda$ is the precision parameter, will be taken unknown and the case of known $\lambda$ will be treated as a special case.  The usual case is that $\bSigma$ is a correlation matrix and $V(\bY) = \lambda^{-1}$ is the common variance for all responses.  However, $\bSigma$ may be any symmetric positive definite matrix, and $\lambda$ is a ``deflation'' parameter.
\end{itemize}
\textbf{Note:} In the Python \verb|pyblm| implementation I tried to use the similar names for variables, e.g. \verb|X| for $\bX$,  \verb|th| for $\btheta$, \verb|la| for $\lambda$ etc.

\medskip
The conjugate prior is a Normal-Gamma, that is, conditional on $\lambda$, $\btheta$ is
$\Nor_p( \btheta_0, \bA_0^{-1})$  and $\lambda$ has a $Ga( \alpha_0, \beta_0)$ distribution.  Namely
$$
\pi( \btheta | \lambda) = (2 \pi)^{-\frac{p}{2}} \lambda^{\frac{p}{2}} \left| \bA_0 \right|^{\frac{1}{2}}
\exp\left\{ -\frac{\lambda}{2} (\btheta - \btheta_0)' \bA_0 (\btheta - \btheta_0) \right\}
$$
and
$$
\pi(\lambda) = \frac{\beta_0^{\alpha_0}}{\Gamma(\alpha_0)} \lambda^{\alpha_0 - 1} \exp\left\{-\lambda \beta_0 \right\}
I_{[0,\infty)}(\lambda),
$$
\begin{itemize}
\item $\btheta_0$ is a $p \times 1$ mean vector for the prior.
\item $\bA_0$ is a $p \times p$ precision matrix for the prior.
\item $\alpha_0 > 0$ and $\beta_0 > 0$ are the \textit{shape}  and \textit{rate} parameters of the Gamma prior for $\lambda$.  In this parametrization the expected value is $\frac{\alpha_0}{\beta_0}$.
\end{itemize}

The model is this, for some responses $\bY = \by$
$$
f( \by | \btheta, \lambda) = (2 \pi)^{-\frac{n}{2}} \lambda^{\frac{n}{2}} \left| \bA \right|^{\frac{1}{2}}
\exp\left\{ -\frac{\lambda}{2} (\by - \bX \btheta)' \bA (\by - \bX \btheta) \right\}
$$
and the posterior is this
\begin{equation}\label{eqn:bayes}
\pi( \btheta, \lambda | \by) = \frac{f( \by | \btheta, \lambda) \pi( \btheta | \lambda) \pi(\lambda)}{f(\by)}
\end{equation}
or, copying everything,
\begin{align}
\pi( \btheta, \lambda | \by) = \label{eqn:post_all1}& 
K (2 \pi)^{-\frac{n+p}{2}} \left| \bA \right|^{\frac{1}{2}} \left| \bA_0 \right|^{\frac{1}{2}} \frac{\beta_0^{\alpha_0}}{\Gamma(\alpha_0)} \lambda^{\frac{n}{2}} \lambda^{\frac{p}{2}} \lambda^{\alpha_0 - 1}\\
&\exp\left\{ -\lambda \left[ \beta_0 ~+~ \frac{1}{2} \left( (\by - \bX \btheta)' \bA (\by - \bX \btheta) ~+~ (\btheta - \btheta_0)' \bA_0 (\btheta - \btheta_0) \right) \right] \right\} \label{eqn:post_all2}, 
\end{align}
where $K = f(\boldsymbol{y})^{-1}$ , the normalizing constant.  The term in the exponential is also a quadratic form, with some constant, and therefore we should be able to write the posterior as
\begin{equation*}
\pi( \btheta, \lambda | \by) \propto
\lambda^{\frac{p}{2}} \exp\left\{ -\frac{\lambda}{2} (\btheta - \btheta_n)' \bA_n (\btheta - \btheta_n) \right\}
\lambda^{\alpha_n - 1} \exp\left\{- \lambda \beta_n \right\}
\end{equation*}
for some mean $\btheta_n$ vector, matrix $\bA_n$ and parameters $\alpha_n = \alpha_0 + \frac{n}{2}$ and $\beta_n$.
If $\bA_n$ is positive definite and $\beta_n$ positive then this is also a Normal-Gamma distribution with parameters  $\btheta_n, \bA_n, \alpha_n, \beta_n$.

\section{The calculations}

In this part I prefer to do the calculations by small steps.  From \eqref{eqn:post_all2} we expand the two quadratic forms

$$
Q = \by' \bA \by ~-~ 2 \by' \bA \bX \btheta ~+~ \btheta' \bX' \bA \bX \btheta ~+~
\btheta' \bA_0 \btheta ~-~ 2 \btheta_0' \bA_0 \btheta ~+~  \btheta_0' \bA_0 \btheta_0 .
$$
which we need to turn into
$$
(\btheta - \btheta_n)' \bA_n (\btheta - \btheta_n) + c =
\btheta' \bA_n \btheta ~-~ 2 \btheta_n' \bA_n \btheta ~+~ \btheta_n' \bA_n \btheta_n + c.
$$
\begin{enumerate}
\item From terms 3 and 4 in $Q$, $\btheta' ( \bX' \bA \bX + \bA_0 ) \btheta = \btheta' \bA_n \btheta$.  Then
$$
\bA_n = \bA_0 +  \bX' \bA \bX .
$$
If $\bX$ is of full rank then $\bX' \bA \bX$ is positive definite and since $\bA_0$ is also positive definite the $\bA_n$ is.  However, even if $\bX$ is not full rank, $\bA_n$ may become positive definite.  For example, a design matrix $\bX$ with colinearity may still be analyzed, using an adequate prior.    

\item From terms 2 and 5 in $Q$, $~-~ 2 ( \by' \bA \bX + \btheta_0' \bA_0 ) \btheta = - 2 \btheta_n' \bA_n \btheta$.   Then
\begin{equation}\label{eqn:th_n}
\bA_n \btheta_n =  \bA_0 \btheta_0 + \bX' \bA \by ~~\text{or}~~
\btheta_n = \bA_n^{-1} ( \bA_0  \btheta_0 + \bX' \bA \by).
\end{equation}
\item From terms 1 and 6 in $Q$, $c = \by' \bA \by + \btheta_0' \bA_0 \btheta_0 - \btheta_n' \bA_n \btheta_n$.
Then \eqref{eqn:post_all2} is
$$
\exp\left\{ -\lambda \left[ \beta_0 + \frac{1}{2} \left( Q + c \right) \right] \right\}
$$
which means that
$$
\beta_n = \beta_0 + \frac{1}{2} \left( \by' A \by + \btheta_0' \bA_0 \btheta_0 - \btheta_n' \bA_n \btheta_n \right) ,
$$
which we need to prove it is always positive.  See Section~\ref{sec:beta_n}.
\end{enumerate}

Following \cite{ohagan1994}, for $\bA = \bI$, let $\hat{\btheta} = (\bX' \bX)^{-1} \bX' \by$, the classic MLE estimator for $\btheta$ (or the Moore-Penrose inverse) and
$$
\bW =  \bA_n^{-1}  \bX' \bX ~~\text{and}~~
\bI - \bW = \bA_n^{-1} (\bA_n - \bX' \bX) = \bA_n^{-1} \bA_0  ,
$$
then
$$
\btheta_n = (\bI - \bW) \btheta_0 + \bW \hat{\btheta} .
$$ 
That is, the posterior mean $\btheta_n$ is a weighted average between the prior mean $\btheta_0$ and $\hat{\btheta}$.

\subsection{Alternative expressions for $\beta_n$}\label{sec:beta_n}

A very helpful shortcut for calculations is this.  Since
\begin{align*}
\bY = & \bX \btheta + \bepsilon; ~ \bepsilon \sim \Nor_n( \bzero, \lambda^{-1} \bA^{-1}) \\
\btheta = &\btheta_0 + \bepsilon_1; ~ \bepsilon_1 \sim \Nor_p( \bzero, \lambda^{-1} \bA_0^{-1}) ,
\end{align*}
then, using basic properties of the Normal, $\bX \btheta$ is Normal distribution and
$$
\bY = \bX \btheta_0 + \bX \bepsilon_1 + \bepsilon
$$ 
and
\begin{equation}\label{eqn:y_marg}
\bY \mid \lambda \sim \Nor_n( \bX \btheta_0, \lambda^{-1} (\bA^{-1} + \bX \bA_0^{-1} \bX' )) .
\end{equation}
The dependency on $\btheta$ has been removed.  The posterior marginal may be calculated directly, since
\begin{align}
\pi( \lambda | \by)\propto &f( \by | \lambda) \pi(\lambda) \nonumber\\
\propto & (2 \pi)^{-\frac{n}{2}} \left| \bA^{-1} + \bX \bA_0^{-1} \bX' \right|^{-\frac{1}{2}}  \frac{\beta_0^{\alpha_0}}{\Gamma(\alpha_0)} \lambda^{\alpha_0 + \frac{n}{2} - 1} \label{eqn:la_y}\\
&\exp\left\{ -\lambda \left[ \beta_0 ~+~ \frac{1}{2} \left( (\by - \bX \btheta_0)' (\bA^{-1} + \bX \bA_0^{-1} \bX' )^{-1} (\by - \bX \btheta_0 ) \right) \right] \right\} . \nonumber
\end{align}
This means that, as already seen,
$$
\lambda | \by \sim Ga( \alpha_n, \beta_n )
$$
with $\alpha_n = \alpha_0 + \frac{n}{2}$ as above and
\begin{equation}\label{eqn:betan_2}
\beta_n = \beta_0 ~+~ \frac{1}{2} \left( (\by - \bX \btheta_0)' (\bA^{-1} + \bX \bA_0^{-1} \bX' )^{-1} (\by - \bX \btheta_0 ) \right) ,
\end{equation}
and we have proved that
$\beta_n = \beta_0 + \frac{1}{2}q( \by | \bX, \btheta_n, \bA_0, \btheta_0)$ where

\begin{align}
\label{eqn:quv}q( \by | \bX, \btheta_n, \bA_0, \btheta_0) =& ~
\by' \bA \by  + \btheta_0' \bA_0 \btheta_0 - \btheta_n' \bA_n \btheta_n \\\notag
=& (\by - \bX \btheta_0)' (\bA^{-1} + \bX \bA_0^{-1} \bX' )^{-1} (\by - \bX \btheta_0 ) .
\end{align}

This means $\beta_n$ is positive if
$(\bA^{-1} + \bX \bA_0^{-1} \bX' )^{-1}$ is positive definite.  The equivalence between the above two expressions may be established by shear matrix algebra and using the Sherman-Woodbury-Morrison lemma, but the equivalence is already proven above.

A third very useful expression for $\beta_n$ is found in the following way.  Note from \eqref{eqn:th_n} that $\bA_n \btheta_n =  \bA_0 \btheta_0 + \bX' \bA \by$ then

\begin{align}
\label{eqn:dspuv}\btheta_n' \bA_n \btheta_n = \btheta_n' \bA_0 \btheta_0 + \by' \bA \bX \btheta_n .
\end{align}
Then, substituting in (\ref{eqn:quv}) and adding and subtracting
$\by' \bA \bX \btheta_n$ and
$\btheta_n' \bX' \bA \bX \btheta_n$, we have,

\begin{align*}
q( \by | \bX, \btheta_n, \bA_0, \btheta_0) =& ~
\by' \bA \by + \btheta_0' \bA_0 \btheta_0 - \btheta_n' \bA_n \btheta_n \\
=& ~ \btheta_0' \bA_0 \btheta_0 + \by' \bA \by - \btheta_n' \bA_0 \btheta_0 - \by' \bA \bX \btheta_n \\
& - \by' \bA \bX \btheta_n + \btheta_n' \bX' \bA \bX \btheta_n + \by' \bA \bX \btheta_n - \btheta_n' \bX' \bA \bX \btheta_n \\
=& ~ (\by - \bX \btheta_n)' \bA ( \by - \bX \btheta_n) ~+ \\
& \btheta_0' \bA_0 \btheta_0 - \btheta_n' \bA_0 \btheta_0 + \by' \bA \bX \btheta_n - \btheta_n' \bX' \bA \bX \btheta_n \\ 
\end{align*}

For the second term we have
\begin{align*}
& \btheta_0' \bA_0 \btheta_0 - \btheta_n' \bA_0 \btheta_0 + \by' \bA \bX \btheta_n - \btheta_n' \bX' \bA \bX \btheta_n \\ 
=& \btheta_0' \bA_0 \btheta_0 - \btheta_0' \bA_0 \btheta_n + \by' \bA \bX \btheta_n - \btheta_n' \bX' \bA \bX \btheta_n 
\end{align*}
\\
Solving for $\by' \bA \bX \btheta_n$ from equation (\ref{eqn:dspuv}) and substituting:
\begin{align*}
    & \btheta_0' \bA_0 \btheta_0 - \btheta_0' \bA_0 \btheta_n + \by' \bA \bX \btheta_n - \btheta_n' \bX' \bA \bX \btheta_n \\ 
=& \btheta_0' \bA_0 \btheta_0 - \btheta_0' \bA_0 \btheta_n + \btheta_n' \bA_n \btheta_n - \btheta_0' \bA_0 \btheta_n - \btheta_n' \bX' \bA \bX \btheta_n \\
=& \btheta_0' \bA_0 \btheta_0 - 2 \btheta_0' \bA_0 \btheta_n + \btheta_n' \bA_0 \btheta_n - \btheta_n' \bA_0 \btheta_n + \btheta_n' \bA_n \btheta_n 
 - \btheta_n' \bX' \bA \bX \btheta_n \\
 =& (\btheta_0 - \btheta_n)' \bA_0 (\btheta_0 - \btheta_n) +
\btheta_n' ( \bA_n - (\bA_0 + \bX' \bA \bX)) \btheta_n, \\
\end{align*}
where indeed the second term in the last equality is zero since $\bA_n = \bA_0 + \bX' \bA \bX$.  Therefore
\begin{equation}\label{eqn:q_y}
q( \by | \bX, \btheta_n, \bA_0, \btheta_0) =
(\by - \bX \btheta_n)' \bA (\by - \bX \btheta_n) +
(\btheta_0 - \btheta_n)' \bA_0 (\btheta_0 - \btheta_n)
\end{equation}
or
$$
\beta_n = \beta_0 ~+~ \frac{1}{2} \left[ (\by - \bX \btheta_n)' \bA (\by - \bX \btheta_n) +
(\btheta_0 - \btheta_n)' \bA_0 (\btheta_0 - \btheta_n)  \right]	,
$$
which is indeed positive.  More importantly,
$(y - \bX \btheta_n)' \bA (y - \bX \btheta_n)$ is the (weighted) sum of the squared residuals and $(\btheta_0 - \btheta_n)' \bA_0 (\btheta_0 - \btheta_n)$ is a kind of penalization for the discrepancy between the prior and posterior means.
Note that
$$
(E[\lambda | \by])^{-1} =
\frac{\beta_n}{\alpha_n} =
\frac{\beta_0 ~+~ \frac{1}{2} \left[ (\by - \bX \btheta_n)' \bA (\by - \bX \btheta_n) +
	(\btheta_0 - \btheta_n)' \bA_0 (\btheta_0 - \btheta_n)  \right]}{\alpha_0 + \frac{n}{2}} .
$$

\section{The normalizing constant, model evidence and the posterior predictive and marginal distributions}\label{sec:nor_cnt}

From \eqref{eqn:bayes} the normalizing constant, which, as a function of $\by$ is the prior predictive model, is
$$
f(\by) = \int f( \by | \lambda) \pi(\lambda) d\lambda .
$$
From \eqref{eqn:la_y} we have
\begin{align}
f(\by) =& (2 \pi)^{-\frac{n}{2}} \left| \bA^{-1} + \bX \bA_0^{-1} \bX' \right|^{-\frac{1}{2}}  \frac{\beta_0^{\alpha_0}}{\Gamma(\alpha_0)} \int \lambda^{\alpha_n - 1} 
\exp\left\{ -\lambda \beta_n \right\} d\lambda \nonumber\\
=&  (2 \pi)^{-\frac{n}{2}}
| \bB_0 |^{\frac{1}{2}}  \frac{\beta_0^{\alpha_0}}{\Gamma(\alpha_0)}
\Gamma(\alpha_n) \beta_n^{-\alpha_n} . \label{eqn:norm_cnst}
\end{align}
where $\bB_0 = ( \bA^{-1} + \bX \bA_0^{-1} \bX' )^{-1}$.  Using the expression for $\beta_n$ in \eqref{eqn:betan_2} it is simple to see that
$$
f(\by) =
\pi^{-\frac{n}{2}} \left| \nu_0^{-1} \bB_0^* \right|^{\frac{1}{2}} \frac{\Gamma\left( \frac{\nu_0 + n}{2}\right) }{\Gamma \left( \frac{\nu_0}{2}\right)}
\left[ 1 + \frac{(\by - \bX \btheta_0)' \bB_0^* (\by - \bX \btheta_0) }{\nu_0}\right]^{-\left(\frac{\nu_0 + n}{2}\right)}
$$
where $\nu_0 = 2 \alpha_0$ and $\bB_0^* = \frac{\alpha_0}{\beta_0} \bB_0$.  This is a non-central multivariate t distribution and therefore, a priori,
$$
\bY \sim T_n^{\nu_0}( \bX \btheta_0, (\bB_0^*)^{-1}) .
$$  
For fixed $\by$ the more compact form of the normalizing constant in \eqref{eqn:norm_cnst} should be used.

\subsection{Model comparisons, variable selection}\label{sec:mod_comp}

In terms of model comparisons, one would have different design matrices $\bX_i$ and possible priors for $\btheta$, of different dimensions, leading to the normalization constants for each model
$f( \by | \bX_i)$; these are called the ``model evidence'' since the posterior probability of model $i$ is \cite{hoeting1999}
$$
\frac{f( \by | \bX_i)}{\sum_{j=1}^{M} f( \by | \bX_j)} .
$$
Assuming that the prior for $\lambda$ remains the same, the model evidence may be taken as 
\begin{equation}\label{eqn:model_evidence}
z( \by | \bX_i) = | \bB_0^i |^{\frac{1}{2}} (\beta_n^i)^{-\alpha_n} ,
\end{equation}
for the obvious definition of $\beta_n^i$ and
$\bB_0^i =  ( \bA^{-1} + \bX_i \bA_0^{-1} \bX_i' )^{-1}$.  It should also be clear that when $\lambda$ is constant we have
$$
z( \by | \bX_i, \lambda) = | \bB_0^i |^{\frac{1}{2}} e^{-\lambda \beta_n^i} .
$$

Using the Sherman-Woodbury-Morrison lemma
it is easy to see that
$$
\bB_0^i =  \bA - \bA \bX_i (\bA_n^i)^{-1} \bX_i' \bA ,
$$
where indeed $\bA_n^i = \bA_0 + \bX_i' \bA \bX_i$.
When we have independent data, i.e. $\bA = \bI$,
$$
\bB_0^i =  \bI - \bX_i (\bA_n^i)^{-1} \bX_i' ,
$$

\subsection{The predictive distribution}\label{sec:pred}

Suppose we want to predict $\bZ$, a $m \times 1$ vector of unobserved responses, corresponding to a $m \times p$ matrix of covariates $\bX_m$, given the data $\by$.  It is more complicated (and interesting) with correlated data.  Using properties of the multivariate Normal, and considering the joint normal distribution of $[ \bY, \bZ]'$, \textbf{conditional on} $\bY = \by$ we have
$$
\bZ | \btheta, \bY=\by, \lambda \sim
\Nor_m( \bX_m \btheta + \bv_m' \bA (\by - \bX \btheta),
\lambda^{-1} ( (\bA^m)^{-1} - \bv_m' \bA \bv_m )   )
$$
where $V(\bZ) = \lambda^{-1}(\bA^m)^{-1}$ the $m \times m$ variance-covariance matrix of $\bZ$ and $cov( \bY, \bZ) = \lambda^{-1} \bv_m$ the $n \times m$ matrix of cross covariances of $[ \bY, \bZ]'$.

Indeed, we already know that
$$
\btheta | \bY=\by, \lambda \sim
\Nor_p ( \btheta_n, \lambda^{-1} \bA_n^{-1})
$$
and therefore
\begin{align*}
& \bZ | \bY=\by, \lambda \sim \Nor_m( \bX_m \btheta_n + \bv_m' \bA (\by - \bX \btheta_n), \\
& \lambda^{-1} ( (\bA^m)^{-1} - \bv_m' \bA \bv_m  + (\bX_m - \bv_m'\bA \bX) \bA_n^{-1} (\bX_m - \bv_m'\bA \bX)')   ).
\end{align*}
(For the conditional mean of $\bZ$ we use the alternative expression $\bv_m'\bA \by + (\bX_m - \bv_m'\bA \bX)\btheta$ to calculate the variance-covariance matrix.)
Note that for independent data $\bA = \bI, \bA_m = \bI$ and $\bv_m = \bzero$ and the expression is far simpler giving
\begin{equation}\label{eqn:pred}
\bZ | \bY=\by, \lambda \sim 
\Nor_m( \bX_m \btheta_n,
\lambda^{-1} ( \bI + \bX_m \bA_n^{-1} \bX_m')   ) .
\end{equation}
These are already the predictive distributions if $\lambda$ is known.  Let
\begin{align*} \label{eqn:t_predc_mu_B}
\bmu_m =& \bX_m \btheta_n + \bv_m' \bA (\by - \bX \btheta_n) \\
\bB_m^{-1} =&
 ( (\bA^m)^{-1} - \bv_m' \bA \bv_m ) + (\bX_m - \bv_m'\bA \bX) \bA_n^{-1} (\bX_m - \bv_m'\bA \bX)'
\end{align*}
then, following the same calculation as in Section~\ref{sec:nor_cnt}, and integrating with respect to the posterior of $\lambda$, we obtain
\begin{equation}\label{eqn:t_pred}
\bZ | \bY=\by \sim T_m^{\nu_m} ( \bmu_m, (\bB_m^*)^{-1}),
\end{equation}
where $\nu_m = 2\alpha_n$ and $\bB_m^* = \frac{\alpha_n}{\beta_n} \bB_m$.

In this fully Bayesian framework, where the predictive distribution is available in closed form, the use of asymptotic model selection criteria such as the BIC is not particularly appropriate. Instead, predictive accuracy provides a more natural basis for model comparison.

\subsection{Posterior marginal distributions}

It should be clear that the marginal posterior for $\lambda$ (indeed, if unknown) is
$$
\lambda \mid \bY = \by \sim Ga( \alpha_n, \beta_n).
$$
To obtain the marginal distribution of the individual parameters $\theta_j$, we prefer to take a more general approach.  Let $\bT$ be any $k \times p$ matrix, then using \ref{property3NM}, we have

$$
\bT \btheta \mid \bY = \by, \lambda \sim
N_k( \bT \btheta_n, \lambda^{-1} \bT \bA_n^{-1} \bT') .
$$
This is already the result for $\lambda$ known.
Using the posterior for $\lambda$ we obtain, as in the previous section,
$$ 
\bT \btheta \mid \bY = \by \sim
T_k^{\nu_m} \left( \bT \btheta_n,
\frac{\beta_n}{\alpha_n} \bT \bA_n^{-1} \bT' \right).
$$
The special case where $\bT = \be_j, \bT \btheta = \theta_j$ is
\begin{equation}\label{eqn:1dmarg}
\theta_j \mid \bY = \by \sim
t_{\nu_n} \left( \theta^n_j,
\frac{\beta_n}{\alpha_n} \sigma_{j,j}^{2}  \right)
\end{equation}
where $\bA_n^{-1} = (\sigma_{i,j}^2)$.

\subsection{Efficient calculations}

Assuming, in the common case, that $n >> p$, then manipulating $\bSigma$ is the most computationally demanding process.
The Cholesky decomposition is performed on the variance-covarince matrix $\bSigma = \bL \bL'$.  One then does Forward Substitution to solve $\bL \bu_i = \be_i$ and these are used as row vectors to construct $\bU$ to obtain $\bA = \bSigma^{-1} = \bU \bU'$ the Cholesky decomposition of the precision matrix (since $\bL \bu_i = \be_i$ then $\bL \bU' = \bI$ or $\bU = (\bL^{-1})'$ and note that $\bA = (\bL \bL')^{-1} = (\bL^{-1})' (\bL^{-1}) = \bU \bU'$).

Note that in calculating $\bA_n$ the unavoidable computational burden is in calculating $\bX' \bA \bX$.  One can then solve $\bA_n \btheta_n = \bA_0 \btheta_0 + \bX' \bA \by$ using LU decomposition, but since this is only a $p \times p$ matrix we might as well simply calculate the inverse.
The latter is in fact a good idea in prediction since $\bA_n^{-1}$ is required, see \eqref{eqn:pred}, for example.  We do the Cholesky and inverse decomposition of $\bA_n$ for all calculations.
Since many calculations involve $\bX' \bA \bX$ it is a good idea first calculate $\bH = \bU \bX$ and use it in all calculations. 

In the case of model comparisons, see \eqref{eqn:model_evidence}, the determinant of $\bB_0^i =  ( \bA^{-1} + \bX_i \bA_0^{-1} \bX_i' )^{-1}$ is required.  Here use the Sherman-Woodbury-Morrison result to see that
$$
\bB_0^i = \bA - \bA \bX_i ( \bA_0 + \bX_i' \bA \bX_i)^{-1} \bX_i' \bA = \bA - \bA \bX_i (\bA_n^i)^{-1} \bX_i' \bA .
$$
From this we may use the matrix-determinant lemma, assuming the Cholesky decomposition of $( \bA_0 + \bX_i' \bA \bX_i)^{-1} = ( \bA_n^i )^{-1}  = \bL_n^i (\bL_n^i)' $ (this is only $p_i \times p_i$)
\begin{align*}
| \bB_0^i | =&
\left| \bA - \bA \bX_i (\bA_n^i)^{-1} \bX_i' \bA  \right| \\
=& \left| \bA - (\bA \bX_i \bL_n^i) (\bA \bX_i \bL_n^i)'  \right| \\
=& |\bA| ( 1 - (\bA \bX_i \bL_n^i)' \bA^{-1} (\bA \bX_i \bL_n^i) ) \\
=& |\bA| ( 1 - (\bL_n^i)' (\bX_i' \bA \bX_i) (\bL_n^i) ) 
 . 
\end{align*}
This should save computational burden since it is no longer required to calculate the determinant of the $n \times n$ matrix $\bB_0^i$.

Finally, in the case of predictive distributions, it is requiered to calculate
$$
\bB_m^{-1} =
( (\bA^m)^{-1} - \bv_m' \bA \bv_m ) + (\bX_m - \bv_m'\bA \bX) \bA_n^{-1} (\bX_m - \bv_m'\bA \bX)'
$$
Using the already available Cholesky decomposition of
$\bA = \bU'\bU$, let $\ba_m = \bU \bv_m$ and $ $
then $\bh_m = (\bX_m - \ba_m' \bU \bX) \bL_n$
$$
\bB_m^{-1} =
( (\bA^m)^{-1} - \ba_m' \ba_m ) + \bh_m \bh_m' .
$$

\section{Particular cases}

\subsection{Regression: formulae}\label{sec:regression}

\begin{figure}
\begin{center}
\includegraphics[scale=0.7]{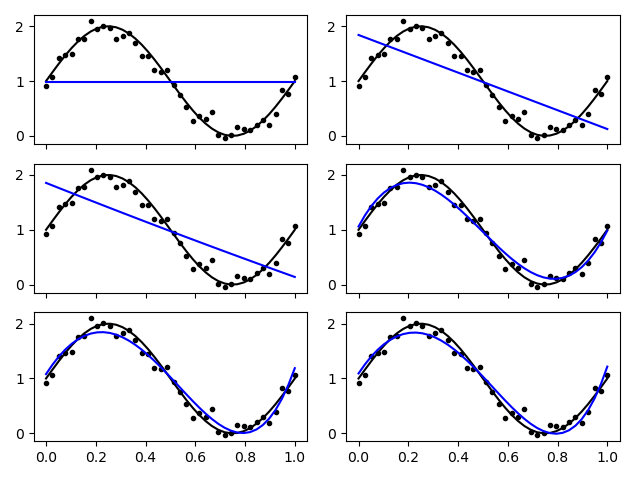}
\caption{\label{fig:Fit} Simulated data and the true function (black curve).  From left to right, top to bottom: MAP fit (blue), that is, the resulting regression using $\btheta_n^p$, for $p=1,2,\ldots,6$.}
\end{center}
\end{figure}

\begin{figure}
\begin{center}
\includegraphics[scale=0.7]{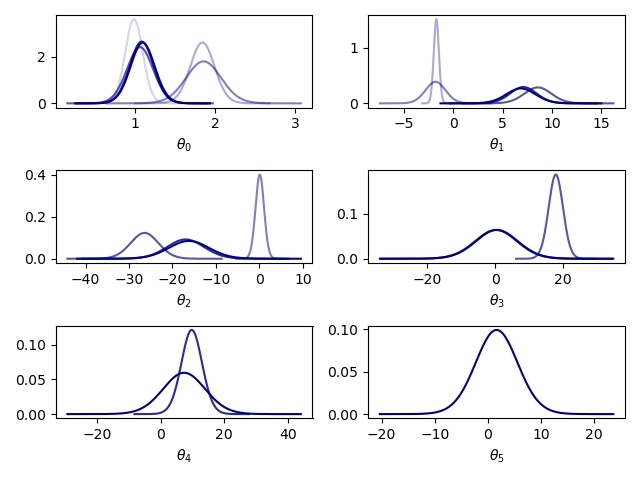}
\caption{\label{fig:Posts} Marginal posterior t distributions for each parameter in the regression, for $p=1,2,\ldots,6$, using \eqref{eqn:1dmarg}.  The marginal posterior for $\lambda \sim Ga( \alpha_n, \beta_n)$ is not included.}
\end{center}
\end{figure}

\begin{figure}
\begin{center}
\includegraphics[scale=0.7]{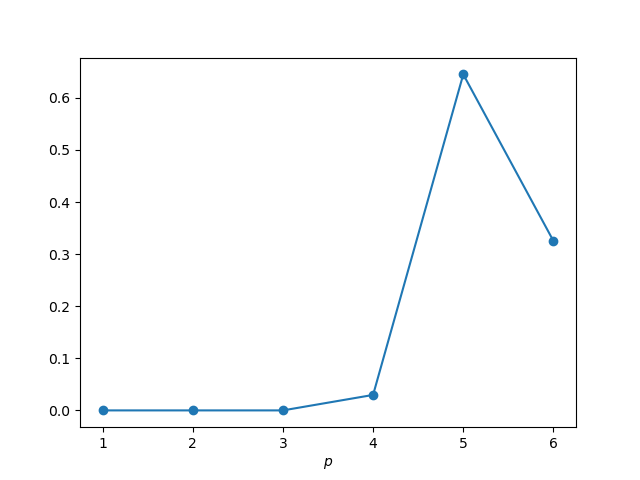}
\caption{\label{fig:ModelPost} Posterior probability of each model, using \eqref{eqn:model_evidence}.}
\end{center}
\end{figure}

The general formulas for regression are outlined in Table~\ref{tab:pars}.
We include an example of a regression, simulating values from
$$
y_i = 1 + \sin( 2 \pi x_i) + \sigma e_i; ~e_i \sim N(0,1),
$$
for $x_i = i/n; i=1,2,\ldots,n$, $n=40$ and $\sigma = 0.1$.  We attempt the regressions
$$
\bY = \bX_p \btheta^p + \epsilon
$$
as in \eqref{eqn:reg} with uncorrelated data, i.e. $\bSigma = \bI$.  The rows of the design matrix are composed with $\bX_p = (\phi_p(x_i))$ with the regressor function $\phi_p(x) = [ 1, x, x^2, \ldots, x^{p-1}]$ and $\btheta^p = [ \theta_0, \theta_1, \ldots, \theta_{p-1}]'$.
For $p=1,2,\ldots,6$ the regressions are performed, also calculating the posterior probability of each model; see Figures~\ref{fig:Fit}, \ref{fig:Posts} and~\ref{fig:ModelPost}. For the prior, $\btheta_0^p = \bzero$ and $\bA_0^p = 0.001 \bI$, that is, the prior precision is 0.1\% of the data precision (diffuse prior).  $\alpha_0 = 1$ and $\beta_0 = 1$.

\begin{table}\label{tab:pars}
\caption{Prior and posterior parameters for the Normal-Gamma conjugate family.  The expressions for $\btheta_n$ and $\bA_n$ are the same when $\lambda$ is a known constant, just remember that in these notes
$\btheta | \lambda \sim \Nor_p( \btheta_0, \lambda^{-1} \bA_0^{-1} )$ and $\btheta \sim | \bY = \by, \lambda \sim \Nor_p( \btheta_n, \lambda^{-1}\bA_n^{-1} ) $ (normally, when $\lambda$ is known, it is not included in the variance-covariance matrix of $\btheta$).  The more common, but equivalent, expression $
\beta_0 + \frac{1}{2} \left( \by' A \by + \btheta_0' \bA_0 \btheta_0 - \btheta_n' \bA_n \btheta_n \right)
$ is commonly used for $\beta_n$.  The one used here is more informative with minimal added computational burden.}
\medspace
\begin{tabular}{| c | c c c |} \hline
&Prior & Posterior & Uncorrelated Data $\bA = \bI$  \\ \hline
&&& \\
$V(\btheta | \lambda)^{-1}$ & $\bA_0$ & $\bA_n = \bA_0 +  \bX' \bA \bX$ & $\bA_0 +  \bX' \bX$  \\ \hline
&&& \\
$E(\btheta | \lambda )$ & $\btheta_0$ & $\btheta_n = \bA_n^{-1} ( \bA_0  \btheta_0 + \bX' \bA \by)$ &
$\bA_n^{-1} ( \bA_0  \btheta_0 + \bX' \by)$  \\ \hline
&&& \\
Discrepancy & & $d^2_n =   (\btheta_0 - \btheta_n)' \bA_0 (\btheta_0 - \btheta_n)$ &  $(\btheta_0 - \btheta_n)' \bA_0 (\btheta_0 - \btheta_n)$ \\ \hline
&&& \\
Res. SS &  & $s^2_n = (\by - \bX \btheta_n)' \bA (\by - \bX \btheta_n)$ & $(\by - \bX \btheta_n)' (\by - \bX \btheta_n)$ \\ \hline
&&& \\
$\lambda$ shape & $\alpha_0$ & $\alpha_n = \alpha_0 + \frac{n}{2}$ & $\alpha_0 + \frac{n}{2}$ \\ \hline
&&& \\
$\lambda$ rate &  $\beta_0$ & $\beta_n = \beta_0 + \frac{1}{2} \left( s^2_n + d^2_n \right)$ & $\beta_0 + \frac{1}{2} \left( s^2_n + d^2_n \right)$ \\ \hline 
\end{tabular}
\end{table}

\subsection{Gaussian processes (GP)}

GP models have many applications and represent a complex topic \cite{rasmussen2006}.  The basics of the Bayesian analysis of GP's is this.  With univariate response, we have a set of observations or measurements (or computer model evaluations)
$y(\bx_i)$ at some points or locations $i=1, 2, \ldots, n$ for $\bx_i \in D$ in some domain.
Let $\phi: D \rightarrow \mathbb{R}^p$ be some regressor function as in Section~\ref{sec:regression}; commonly
$\phi(x) = [ 1, x_1, x_2, \ldots, x_p]$, the point coordinates.
Let the design matrix be $\bX = (\phi(\bx_i))$ and the response vector $\by = (y(\bx_i))$

\begin{figure}
\begin{center}
\begin{tabular}{c c}
\includegraphics[scale=0.5]{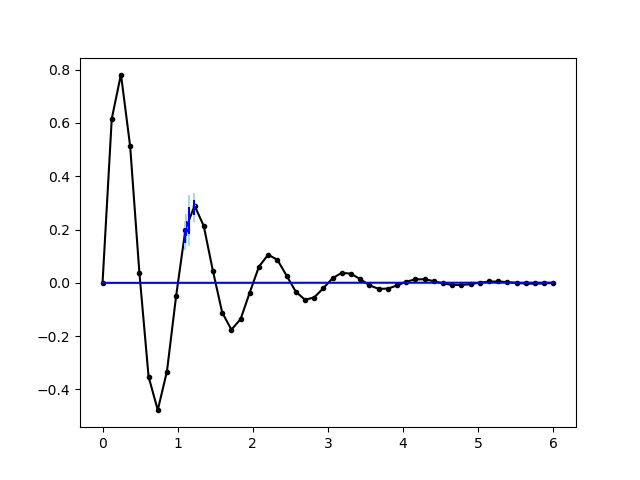} &
\includegraphics[scale=0.5]{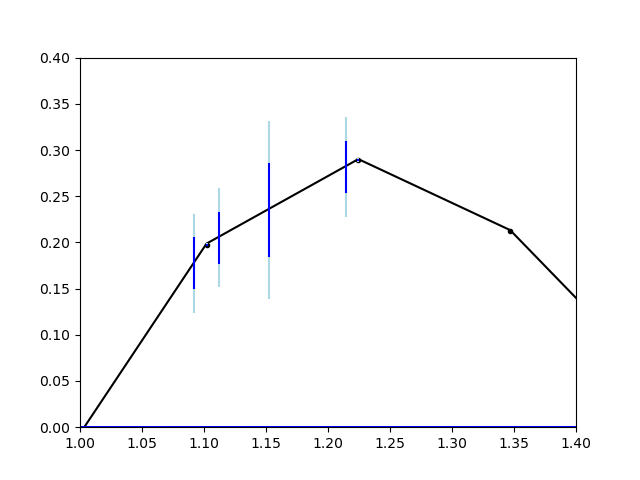}
\end{tabular}
\caption{\label{fig:GP} Gaussian process interpolator; examples of the t predictive distribution (depicted with quantiles) at various locations.  The variance becomes zero at observations points and $\mu(\bx_i) = y(\bx_i)$ (interpolator).}
\end{center}
\end{figure}

A very important aspect of GP is its covariance structure (indeed $\bA \neq \bI$).  For stationary isotropic covariance the covariance is assumed as $cov( y(\bx_i), y(\bx_j)) = \lambda^{-1} k(d(\bx_i, \bx_j))$, where $k$ is a positive defined correlation function \cite{rasmussen2006}; $d$ is a metric in $D$.  From this, the covariance matrix $\bSigma$ is formed.  The idea is to use the predictive distribution $y(\bx)$ at any location $\bx \in D$ as a predictor of the underlying process or function.  One uses the Normal predictive \eqref{eqn:pred}, if $\lambda$ is known, or otherwise the t predictive \eqref{eqn:t_pred}.  Note that in any case, translating to this context, the mean is
$$
\bmu(\bx) =  \phi(\bx) \btheta_n + \bv(\bx)' \bA (\by - \bX \btheta_n) ,
$$
where $\bv(\bx)' = [k(d( \bx, \bx_i))]$.   If $k$ is continuous at zero, the predictor is an interpolator, that is $\mu(\bx_i) = y(\bx_i)$ with zero variance; see Figure~\ref{fig:GP}. 

To build a smoother one includes a discontinuity of $k$ at zero (in geostatistics this is called ``the nugget'', a measuring noise for $y$).For example
$cov( y(\bx_i), y(\bx_j)) = \lambda^{-1} ( k( \bx_i, \bx_j) + \delta_{i,j} \sigma^2)$.

\subsection{Bayesian Dynamic Linear Models (DLM)}

\cite{west1997} is the standard reference in this case.
The general DLM is explained in section 4.2, and the univariate DLM in p. 109 of \cite{west1997}.  The latter is
\begin{align*}
Y_t &= \phi_t \btheta_t + \epsilon;  \epsilon \sim \Nor( 0, \lambda^{-1}) ~~\text{(observation equation)} \\
\btheta_t &= \bsy{G}_t \btheta_{t-1} + \bepsilon_1; \bepsilon_1 \sim \Nor_p( \bzero, \lambda^{-1} (\bA_{t-1})^{-1}) ~~\text{(system equation)},
\end{align*}
with some prior $\btheta_0 | \lambda_0 \sim \Nor_p( \bmu_0, \lambda^{-1} (\bA_0)^{-1})$.

For calculating the posterior distributions the strategy is the following: when observing the first $Y_1 = y_1$, the prior (given $\lambda$) for its parameter is $\btheta_1 \sim \Nor_p( \bsy{G}_1 \bmu_0, \bsy{G}_1 \lambda^{-1} (\bA_0)^{-1} \bsy{G}_1')$, in which case we see that the posterior is $\btheta_1 | Y_1=y_1 \sim
N_p( \bmu_1, \lambda^{-1} (\bA_1)^{-1})$, using the corresponding formulas in Table~\ref{tab:pars}.  We can then repeat the process, using this a prior to obtain the posterior $\btheta_2 | Y_1=y_1, Y_2=y_2$ etc.  These become the ``updating equations'' as seen in theorem 4.1 of \cite{west1997}.  Moreover, one can calculate the predictive distribution of $Y_{t+1} | \bY_t=\by_t$ with the same formulas in Section~\ref{sec:pred}.

\section{Final comments}

Other simpler models may be deduced from these general calculations, as for example standard normal sampling, setting $p=1$ and $X$ a column vector of 1's.  This last model may be used a null model in regression, and use the Bayesian model comparison in Section~\ref{sec:mod_comp}.  

I created a Python class, BLM, to do all the calculations and an example file with the two examples in Regression and in GP's, precisely the examples presented here:
\url{https://github.com/andreschristen/BLM} 

\bibliography{BLM}

\end{document}